\documentstyle[11pt,epsf]{article}
 1
 1
 1

\def\lsim{\mathrel{\raise.3ex\hbox{$<$\kern-.75em\lower1ex\hbox{$\sim$}}}}
\def\gsim{\mathrel{\raise.3ex\hbox{$>$\kern-.75em\lower1ex\hbox{$\sim$}}}}
\catcode`\@=11
\def\slash{\mathpalette\make@slash}
\def\make@slash#1#2{\setbox\z@\hbox{$#1#2$}%
  \hbox to 0pt{\hss$#1/$\hss\kern-\wd0}\box0}
\catcode`\@=12 
\begin{document}
\noindent
\thispagestyle{empty}
\renewcommand{\thefootnote}{\fnsymbol{footnote}}
\begin{flushright}
{\bf UCSD/PTH 97-39}\\
{\bf hep-ph/9801275}\\
{\bf January 1998}\\
\end{flushright}
\vspace{.5cm}
\begin{center}
  \begin{Large}\bf
Top Quark Pair Production at Higher Energies \\[1mm]
at the First Muon Collider
\footnote{Talk presented at the Workshop on Physics at the
First Muon Collider and the Front End of a Muon Collider, Fermilab,
November 6 - 9, 1997, to appear in the proceedings.}
%
  \end{Large}
  \vspace{1.5cm}

\begin{large}
 A.H. Hoang
\end{large}
\begin{center}
\begin{it}
   Department of Physics,
   University of California, San Diego,\\
   La Jolla, CA 92093--0319, USA\\ 
\end{it} 
\end{center}

  \vspace{1.5cm}
  {\bf Abstract}\\
\vspace{0.3cm}
%
\noindent
\begin{minipage}{15.0cm}
\begin{small}
For a proper theoretical description of the top quark pair production
cross section at the First Muon Collider in the kinematic regime away
from the threshold the complete mass and energy dependence has to be
taken into account. Two-loop calculations for the photon mediated
total cross section are reviewed.
\end{small}
\end{minipage}
\end{center}
\setcounter{footnote}{0}
\renewcommand{\thefootnote}{\arabic{footnote}}
\vspace{1.2cm}
\noindent
For experiments at the $Z$ pole quark mass effects in the total
hadronic cross section $\sigma(e^+e^-\to {\rm hadrons})$ can often be
neglected in the first approximation and treated perturbatively in an
expansion in the ratio $M_{\rm quark}^2/s$, where $\sqrt{s}$ is the
c.m.~energy. The situation is quite different for top quark pair
production at the First Muon
Collider, where c.m.~energies up to $500$~GeV are planned. Here, the
$t\bar t$ threshold energy\footnote{
Throughout this talk $M_t$ is understood as the top quark pole mass.
The constants $C_A=3$, $C_F=4/3$ and $T=1/2$ denote SU(3) group
theoretical factors and $\alpha$ is the electromagnetic coupling.
}
$(\sqrt{s})_{thr} = 2 M_t\approx 350$~GeV and the c.m.\ energies are
comparable in size and an expansion in $M_t^2/s$ is not adequate.
In the immediate vicinity of the threshold point conventional
multi-loop perturbation theory breaks down and a resummation of
Coulomb terms $\sim (C_F\,\alpha_s\,\pi/\beta)^n$, $n=0,1,2,\ldots$ 
($\beta = (1-4M_t^2/s)^{1/2}$ being the c.m.\ velocity of the top
quarks), has to be
carried out to all orders in the strong coupling 
(see~\cite{Berger1,Hoang1}). This complication is avoided if only
energies sufficiently above the threshold point are considered. As a
rule of thumb one can take the relation $C_F \alpha_s \pi/\beta < 2$
as a requirement to avoid the complication of the threshold
regime~\cite{CHKST3}. For $t\bar t$ production this means that
conventional multi-loop methods can be used for c.m.\ energies which
are at least $10$~GeV above the threshold point. 

\begin{figure}[htb] 
\begin{center}
\leavevmode
\epsfxsize=4cm
\epsffile[220 420 420 550]{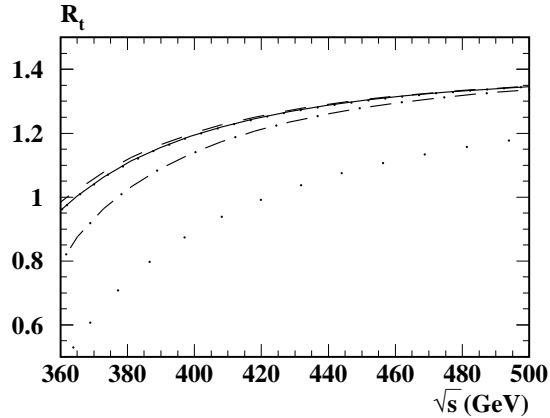}\\
%
%
\vskip  3.0cm
 \caption{\label{fig1} 
The total normalized photon-mediated cross section at the two-loop
level versus $\sqrt{s}$ for the renormalization scales $\mu=M_t$
(dashed), $\mu=2 M_t$ (solid) and $\mu=\sqrt{s}$ (dotted line),
$M_t=175$~GeV and $\alpha_s^{(5)}(M_z)=0.118$. For
comparison also the Born (wide dots) and the one-loop cross section
for $\mu=2 M_t$ (dashed-dotted line) are displayed. To improve the
stability under renormalization scale variations the known three-loop
${\cal{O}}(\alpha_s^3)$ in the large momentum
expansion have been added to the two-loop cross section.
The plot is taken from~[3].
}
 \end{center}
\end{figure}

\begin{figure}[htb] 
\begin{center}
\leavevmode
\epsfxsize=4cm
\epsffile[220 420 420 550]{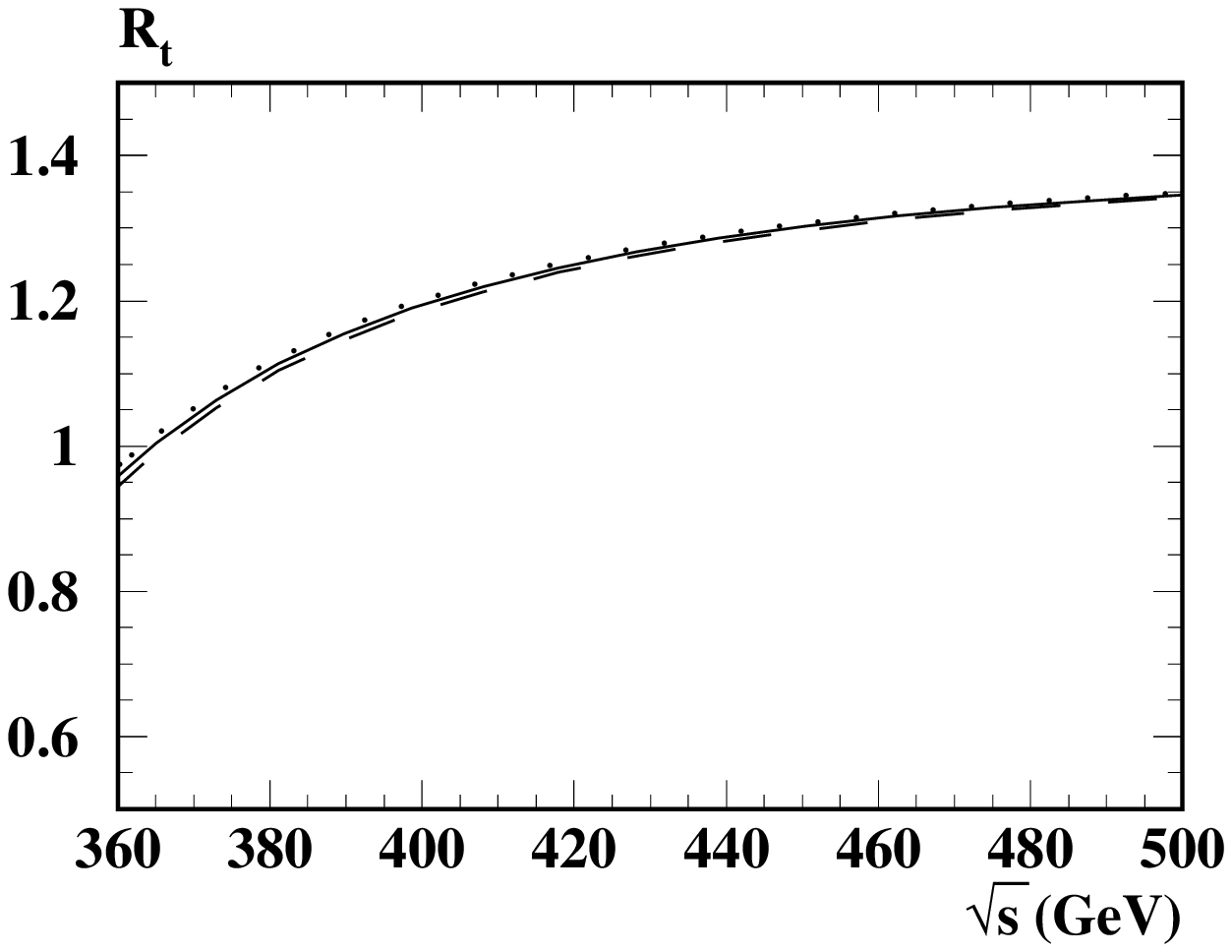}\\
\vskip 0.8cm
\mbox{\hspace{3.5cm}}(a)
\vskip 1.8cm
\leavevmode
\epsfxsize=4cm
\epsffile[220 420 420 550]{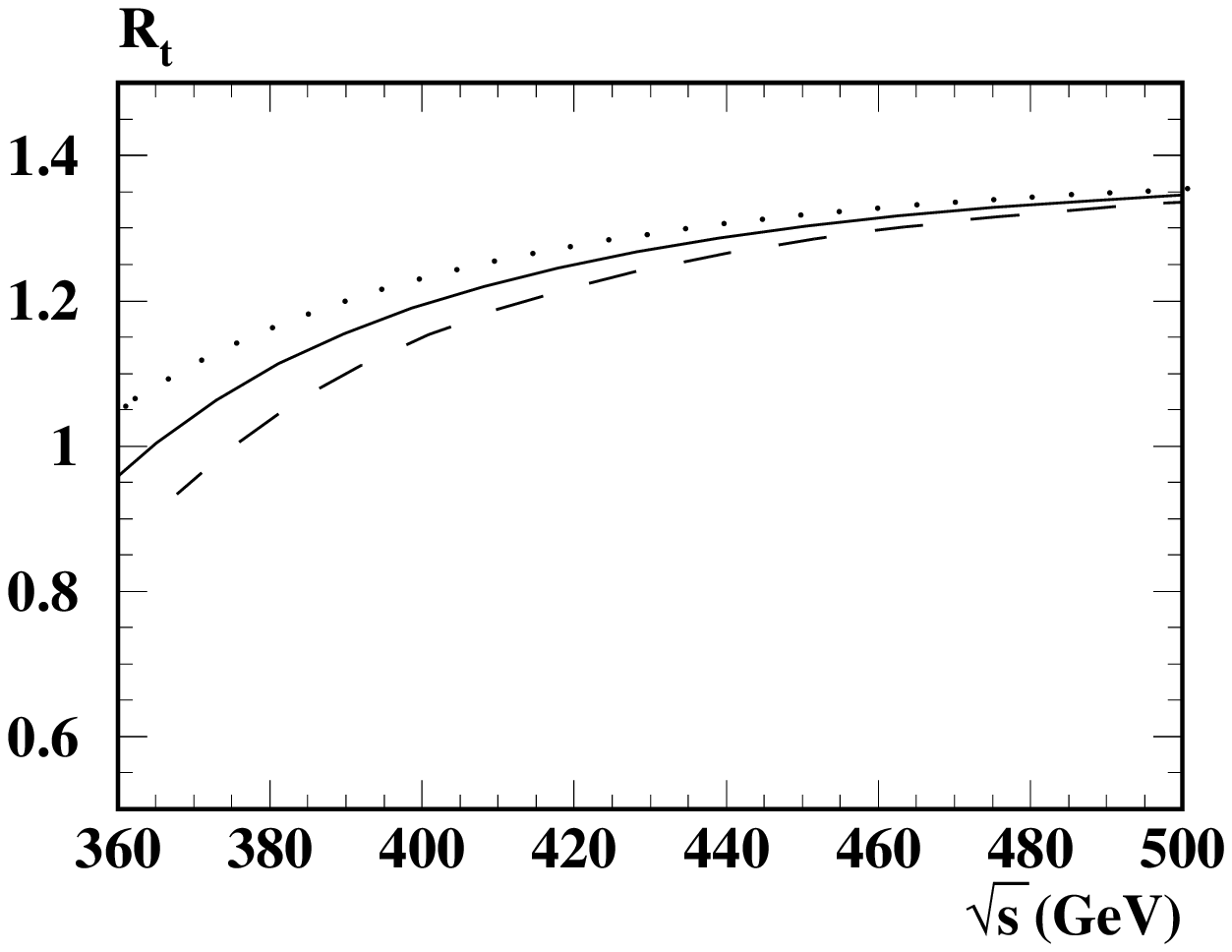}\\
\vskip 0.8cm
\mbox{\hspace{3.5cm}}(b)
%
%
\vskip  1.5cm
 \caption{\label{fig2} 
Sensitivity of the total normalized photon-mediated cross section at
the two-loop level on variations of the strong coupling and the top
quark mass. In (a) the solid, dashed and dotted lines correspond to
$\alpha_s^{(5)}(M_z)=0.118, 0.115$ and $0.121$, respectively, for
$M_t=175$~GeV. In (b) the solid, dashed and dotted lines correspond to
$M_t=175, 180$ and $170$~GeV, respectively, for
$\alpha_s^{(5)}(M_z)=0.118$.
For both plots the choice $\mu=2 M_t$ has been made. The plots are taken
from~[3].
}
 \end{center}
\end{figure}

In this talk I report on developments in the past two years in the
determination of two-loop ${\cal{O}}(\alpha_s^2)$ QCD corrections to
the total $t\bar t$
production cross section where the full dependence on the top quark
mass has been taken into account. Since the results for the production
of $t\bar t$ pairs through the axial-vector current are not completed
yet,\footnote{
Two-loop corrections for the total Z-mediated production cross section
have been calculated recently in an expansion in $M_t^2/s$ up to order
$(M_t^2/s)^6$~\cite{Harlander1}. These results, however, have only
limited applicability for the First Muon Collider, in particular for
lower energies~\cite{CHKS1}.  
}
only the photon-mediated contributions are considered. Therefore,
the result presented here cannot be considered as absolute predictions
for $t\bar t$ production at the First Muon Collider. Rather they
should be taken as an indication for the size and importance of all the
${\cal{O}}(\alpha_s^2)$ contributions. In particular, they can
illustrate the stability of the results for variations in the
renormizalization scale and their sensitivity on the strong coupling and
the top quark mass. Assuming the notation and conventions used
in~\cite{CHKST3}, the total photon-mediated $t\bar t$ cross section
normalized to the point cross section $\sigma_{\rm pt}=4 \pi
\alpha^2/3 s$ can be cast into the form,
\begin{eqnarray}
\lefteqn{
R \, = \, \frac{\sigma(e^+e^-\to\gamma^*\to t\bar
   t)}{\sigma_{\rm pt}} \, = \,
  Q_t^2\,\bigg[\,
  R^{(0)} 
  + \bigg(\frac{\alpha_s^{(5)}(\mu^2)}
         {\pi}\bigg)\,C_F\,R^{(1)}
}
\nonumber\\[2mm] & & \qquad
 + \bigg(\frac{\alpha_s^{(5)}(\mu^2)}{\pi}\bigg)^2\,
 \bigg(\,C_F^2\,R_A^{(2)} + C_F\,C_A\,R_{NA}^{(2)} +
        C_F\,T\,n_l\,R_l^{(2)} + C_F\,T\,R_F^{(2)}
 \,\bigg)
 \,\bigg]
\,.
\end{eqnarray}
$R^{(0)}$ an $R^{(1)}$ denote the well known expressions for the Born
cross section and the one-loop~\cite{Kallensabry1} corrections.
$\alpha_s(\mu^2)$ is the strong coupling in the $\overline{\mbox{MS}}$
scheme at the scale $\mu$ for $n_l=5$ light flavors. 
The Abelian and non-Abelian parts $R_A^{(2)}$
and $R_{NA}^{(2)}$ come from two gluon exchange two-loop diagrams and
have been calculated numerically in~\cite{CKS1,CKS2,CKS3} using Pad\`e
approximation methods. The uncertainties coming from possible choices
of different Pad\`e approximants can be ignored for practical
purposes. Handy approximation formulae for $R_A^{(2)}$
and $R_{NA}^{(2)}$ which reduce to the correct analytic expressions in
the high energy and the small velocity limit can be found
in~\cite{CKS2}. $R_l^{(2)}$ and $R_F^{(2)}$
originate from one gluon exchange diagrams with the insertion of a
massless and a top quark loop, respectively. $R_l^{(2)}$ also contains
the effects from real radiation of a massless quark pair through gluon
radiation off one of the top quarks. It should be noted that both, the
real and virtual radiation of massless quark pairs, have to be taken
into account to render the ${\cal{O}}(\alpha_s^2)$ contributions in
$R_l^{(2)}$ infrared finite. $R_l^{(2)}$ and $R_F^{(2)}$ have been
calculated in closed analytic form in~\cite{HKT1,HT1} using dispersion
relation techniques for the virtual corrections and integration of the
four-body phase space for the real corrections. In~\cite{CKS3} and
\cite{HT1} also the complete set of ``non-singlet'' axial-vector
current induced ${\cal{O}}(\alpha_s^2)$ corrections can be found.
The ${\cal{O}}(\alpha_s^2)$ contributions coming from secondary
radiation of $t\bar t$ pairs through gluon radiation off massless
quarks are numerically small and can be neglected~\cite{CHKST2}.
To fix our notation explicitly I present the complete form of the Born
and one-loop contributions,
\begin{eqnarray}
R^{(0)} & = & \frac{3}{2}\,\beta\,(3-\beta^2)\,,\nonumber\\[2mm]
R^{(1)} & = & 3\,\bigg\{\,
\frac{\left( 3 - {\beta^2} \right) \,\left( 1 + {\beta^2} \right) }{2
    }
\,\bigg[\, 2\,\mbox{Li}_2(p) + \mbox{Li}_2({p^2}) + 
     \ln p\,\Big( 2\,\ln(1 - p) + \ln(1 + p) \Big) 
      \,\bigg] \,\nonumber\,\\[2mm]
 & & \mbox{}\quad - 
  \beta\,( 3 - {\beta^2} ) \,
   \Big( 2\,\ln(1 - p) + \ln(1 + p) \Big)  
\nonumber\,\\[2mm]
 & & \mbox{}\quad 
  - \frac{\left( 1 - \beta \right) \,
     \left( 33 - 39\,\beta - 17\,{\beta^2} + 7\,{\beta^3} \right) }{16}\,
   \ln p\,
  + \frac{3\,\beta\,\left( 5 - 3\,{\beta^2} \right) }{8}
\,\bigg\}
\,,
\end{eqnarray}
where $p=(1-\beta)/(1+\beta)$.
In Fig.~\ref{fig1} R is plotted versus $\sqrt{s}$ including
successively higher orders in $\alpha_s$. The ${\cal{O}}(\alpha_s^2)$
cross section is displayed for different choices of the
renormalization scale. The stability under renormalization scale
variations has been improved by also including the known three-loop
${\cal{O}}(\alpha_s^3)$ in the large momentum
expansion~\cite{GorKatLar91SurSam91,CheKue90,CheKue97}. 
In Fig.~\ref{fig2}(a) the sensitivity of R on the strong coupling and
in Fig.~\ref{fig2}(b) on the top quark mass are illustrated. \\[2mm]
This work is supported in part by
the U.S.~Department of Energy under contract
No.~DOE~DE-FG03-90ER40546. 
%
%

\sloppy
\raggedright
\def\app#1#2#3{{\it Act. Phys. Pol. }{\bf B #1} (#2) #3}
\def\apa#1#2#3{{\it Act. Phys. Austr.}{\bf #1} (#2) #3}
\def\lhc{Proc. LHC Workshop, CERN 90-10}
\def\npb#1#2#3{{\it Nucl. Phys. }{\bf B #1} (#2) #3}
\def\nP#1#2#3{{\it Nucl. Phys. }{\bf #1} (#2) #3}
\def\plb#1#2#3{{\it Phys. Lett. }{\bf B #1} (#2) #3}
\def\prd#1#2#3{{\it Phys. Rev. }{\bf D #1} (#2) #3}
\def\pra#1#2#3{{\it Phys. Rev. }{\bf A #1} (#2) #3}
\def\pR#1#2#3{{\it Phys. Rev. }{\bf #1} (#2) #3}
\def\prl#1#2#3{{\it Phys. Rev. Lett. }{\bf #1} (#2) #3}
\def\prc#1#2#3{{\it Phys. Reports }{\bf #1} (#2) #3}
\def\cpc#1#2#3{{\it Comp. Phys. Commun. }{\bf #1} (#2) #3}
\def\nim#1#2#3{{\it Nucl. Inst. Meth. }{\bf #1} (#2) #3}
\def\pr#1#2#3{{\it Phys. Reports }{\bf #1} (#2) #3}
\def\sovnp#1#2#3{{\it Sov. J. Nucl. Phys. }{\bf #1} (#2) #3}
\def\sovpJ#1#2#3{{\it Sov. Phys. LETP Lett. }{\bf #1} (#2) #3}
\def\jl#1#2#3{{\it JETP Lett. }{\bf #1} (#2) #3}
\def\jet#1#2#3{{\it JETP Lett. }{\bf #1} (#2) #3}
\def\zpc#1#2#3{{\it Z. Phys. }{\bf C #1} (#2) #3}
\def\ptp#1#2#3{{\it Prog.~Theor.~Phys.~}{\bf #1} (#2) #3}
\def\nca#1#2#3{{\it Nuovo~Cim.~}{\bf #1A} (#2) #3}
\def\ap#1#2#3{{\it Ann. Phys. }{\bf #1} (#2) #3}
\def\hpa#1#2#3{{\it Helv. Phys. Acta }{\bf #1} (#2) #3}
\def\ijmpA#1#2#3{{\it Int. J. Mod. Phys. }{\bf A #1} (#2) #3}
\def\ZETF#1#2#3{{\it Zh. Eksp. Teor. Fiz. }{\bf #1} (#2) #3}
\def\jmp#1#2#3{{\it J. Math. Phys. }{\bf #1} (#2) #3}
\def\yf#1#2#3{{\it Yad. Fiz. }{\bf #1} (#2) #3}

\end{document}